# Avoiding the Internet of Insecure Industrial Things

Dr Lachlan Urquhart, Research Fellow in Information Technology Law, Horizon, University of Nottingham.[1]

Professor Derek McAuley, Director of Horizon and Professor of Digital Economy, Horizon, University of Nottingham.

**Abstract:**
Security incidents such as targeted distributed denial of service (DDoS) attacks on power grids and hacking of factory industrial control systems (ICS) are on the increase. This paper unpacks where emerging security risks lie for the *industrial internet of things*, drawing on both technical and regulatory perspectives. Legal changes are being ushered by the European Union (EU) Network and Information Security (NIS) Directive 2016 and the General Data Protection Regulation 2016 (GDPR) (both to be enforced from May 2018). We use the case study of the emergent smart energy supply chain to frame, scope out and consolidate the breadth of security concerns at play, and the regulatory responses. We argue the industrial IoT brings four security concerns to the fore, namely: appreciating the shift from offline to online infrastructure; managing temporal dimensions of security; addressing the implementation gap for best practice; and engaging with infrastructural complexity. Our goal is to surface risks and foster dialogue to avoid the emergence of an *Internet of Insecure Industrial Things*.

## 1. Introduction to the Industrial IoT

The industrial internet of things (IIoT) is an emerging commercial trend that seeks to improve management of the creation, movement and consumption of goods and services. It is part of a wider shift towards cyber physical systems (CPS) which are "…*integrations of computation with physical processes…embedded computers and networks monitor and control the physical processes, usually with feedback loops where physical processes affect computations and vice versa…*".[2] IIoT is distinct from the consumer led IoT trend where ambient sensing occurs by remotely controllable and constantly connected physical objects embedded in domestic settings. These devices with a digital presence and backend computational infrastructure (e.g. cloud, databases, servers), networking and an associated ecosystem of stakeholders[3]. The IIoT departs by applying these technologies to industrial contexts. Instead of convenience, comfort or entertainment, the goal is to increase connectivity and track activity across supply chains.

IIoT is set for significant growth, estimated by Accenture to add $14.2 trillion to the global economy by 2023.[4] Major industrial investment in manufacturing, energy and transportation[5]

---

[1] Corresponding Author Email Address:  lachlan.urquhart@nottingham.ac.uk;
Physical Address: Horizon Digital Economy Research Institute, University of Nottingham Innovation Park, Triumph Road, Nottingham, NG7 2TU.
[2] Edward A Lee, "Cyber Physical Systems: Design Challenges," *Technical Report No. UCB/EECS-2008-8*, 2008, http://www.eecs.berkeley.edu/Pubs/TechRpts/2008/EECS-2008-8.html.
[3] Lachlan Urquhart and Tom Rodden, "New Directions in Information Technology Law: Learning from Human–computer Interaction," *International Review of Law, Computers & Technology* 31, no. 2 (2017): 1–19. – their working definition is derived from surveying a range of IoT stakeholder definitions e.g. Internet Engineering Task Force; International Telecommunications Union; Cisco; Internet Society etc.
[4] Accenture Technology, "Driving Unconventional Growth through the Industrial Internet of Things," 2015, https://www.accenture.com/gb-en/_acnmedia/Accenture/next-gen/reassembling-industry/pdf/Accenture-Driving-Unconventional-Growth-through-IIoT.pdf.
[5] World Economic Forum / Accenture, "Industrial Internet of Things: Unleashing the Potential of Connected Products and Services" (Cologny, 2015), http://www3.weforum.org/docs/WEFUSA_IndustrialInternet_Report2015.pdf.

is in automation, data driven sensing and actuation.[6] In a review of the domain, Xu et al highlight the following key use cases:

- *Healthcare services* - tracking healthcare inventory, global access and sharing of health data, and personalisation of patient care.
- *Food supply chains* - monitoring production from farm to plate including provenance tracking through Radio Frequency ID (RFID), distributed infrastructure and networking.
- *Mining* - safety applications like early warning sensing for natural disasters, chemical and biological sensors for worker disease detection underground.
- *Transport and logistics* - tracking physical objects being transported from origin to destination.
- *Firefighting* - detecting possible fires and providing early warning.[7]

Given the ubiquity of possible IIoT contexts, the breadth of risks can be vast, especially when intersecting with consumer led IoT.[8] For IIoT in healthcare, hacking of insulin pumps or pacemakers is a noteworthy concern.[9] Similarly, in the food supply chain, use of agricultural drones to survey farmland raises concerns for drone hacking, especially for larger vehicles.[10] More broadly though, the industrial threat landscape already involves a multitude of actors utilising different IT vulnerabilities to leverage a variety of attacks.[11] These include:

- State sponsored hackers attacking foreign infrastructure either in advanced persistent threats (APTs) to steal military secrets or intelligence, or in patriotic campaigns to spread propaganda and interfere with foreign elections.[12] APTs often use zero day vulnerabilities (unpatched security flaws) in software to compromise critical infrastructure and steal confidential information.[13] There can also be **commercial cyber-espionage and sabotage** to obtain commercial intelligence, gain competitive advantage over rival businesses, and cause down-time.[14]

- Organised criminal groups **hacking** into organisations to access compromising information (e.g. trade secrets, emerging intellectual property, evidence of malpractice).[15] They may also use malware campaigns to infect laptops or smartphones with remote access tools to record victims on their webcams in precarious acts and extorting them to prevent release of the footage as part of ransomware campaigns.[16]

- Loosely united hacker collective groups, like Lulzsec or Anonymous, use hacking or DDoS attacks[17] for social justice and retaliation against organisations for perceived immoral acts.[18] They will target websites or critical infrastructure to create service disruption and downtime, with associated financial and reputational costs.[19]

---

[6] Li Da Xu et al., "Internet of Things in Industries: A Survey," *IEEE Transactions on Industrial Informatics* 10, no. 4 (2014), doi:10.1109/TII.2014.2300753.
[7] Ibid.
[8] Derek O'Halloran and Elena Kvochko, "Industrial Internet of Things : Unleashing the Potential of Connected Products and Services," *World Economic Forum*, no. January (2015): 40.
[9] Iain Thomson, "BBC's Micro:bit Turns out to Be an Excellent Drone Hijacking Tool • The Register," *The Register*, 2017, https://www.theregister.co.uk/2017/07/29/bbcs_microbit_drone_hijacking_tool/.
[10] Jim Finkle, "J & J Warns Diabetic Patients : Insulin Pump Vulnerable to Hacking," *Reuters*, 2016, http://uk.reuters.com/article/us-johnson-johnson-cyber-insulin-pumps-e-idUKKCN12411L.Lily Hay Newman, 'Medical Devices Are the Next Security Nightmare', *Wired*, 2017, https://www.wired.com/2017/03/medical-devices-next-security-nightmare/
[11] ENISA, *Threat Landscape Report 2016* (ENISA, Heraklion, 2017), 67–72.
[12] Dmitri Alperovitch, "Revealed: Operation Shady RAT," *White Paper*, 2011, https://www.mcafee.com/us/resources/white-papers/wp-operation-shady-rat.pdf.
[13] Brendan Koerner, "Inside the OPM Hack, The Cyberattack That Shocked the US Government," *Wired*, 2016, https://www.wired.com/2016/10/inside-cyberattack-shocked-us-government/.
[14] Thomas Rid, *Cyber War Will Not Take Place* (Hurst & Company, 2013). ; German Steel Mill example, discussed in more detail below
[15] Marisa Randazzo et al., "Insider Threat Study: Illicit Cyber Activity in the Banking and Finance Sector," *Software Engineering Institute*, June 1, 2005, http://repository.cmu.edu/sei/457.
[16] Rebecca S. Portnoff et al., "Somebody's Watching Me?: Assessing the Effectiveness of Webcam Indicator Lights," *Proceedings of the ACM CHI'15 Conference on Human Factors in Computing Systems* 1 (2015): 1649–58, doi:10.1145/2702123.2702164.
[17] Distributed Denial of Service

- Individuals can also create disruption. **Insider threats** posed by disgruntled employees involve use of their internal system access and credentials, or 'social engineering' attacks, to get sensitive information that can be traded with the highest bidder.[20] **Solitary** hackers breaking into military or national security infrastructure from their bedroom seeking to prove existence of UFOs or similar sometimes grab headlines as possible threats but ultimately spend years fighting unbending extradition processes.[21]

The advent of IIoT means vulnerabilities are becoming harder to detect and secure as systems go online. Sadeghi et al argue IIoT security is challenging because security countermeasures will develop slowly (often only prompted in the wake of attacks), the breadth of attack surfaces are wide (e.g. hardware, software, communication protocols etc.) and scope for system failures causing harm to property or humans is significant.[22]

The wealth of stakeholders operating in this domain is another practical issue. Large legitimate and illegitimate cybersecurity economies encapsulate security vendors, consultants and IT firms trying to patch or address threats contrasted with threat agents finding, stockpiling and trading vulnerabilities.[23] This diversity of actors can create confusion. The label 'hacker' is a useful example. Simply put, hackers can sit on a spectrum from law abiding 'white hats' to criminal 'black hats', with 'grey hats' sitting between the two. However, as we see above, it can include organised crime groups, state supported bodies and lone hackers, to name a few.

Weber argues that the only constant in cybersecurity is change, but that it is regulated in a fragmented manner.[24] He argues multiple stakeholders, particularly industry (who are most familiar with issues) and a breadth of regulatory mechanisms are needed to regulate IIoT.[25] Top down state centric legal approaches alone will not suffice.[26] In the privacy domain, we've argued the important role of non-state actors' practices in regulation, and the use of design orientated approaches to tackle regulatory harms from IoT.[27] Despite these challenges, the tide remains against IoT specific legislation in both US and EU, primarily due to desire to give the nascent industry a chance to sort itself out[28] instead favouring industry self-regulation or use of existing law.[29]

In practice whilst we see multi-stakeholder governance against cybersecurity harms, from regional laws to industry standards and initiatives, criminalisation by individual states remains

---

[18] Pammy Olson, *We Are Anonymous : Inside the Hacker World of LulzSec, Anonymous, and the Global Cyber Insurgency* (Back Bay Books 2013).
[19] Argyro P. Karanasiou, "The Changing Face of Protests in the Digital Age: On Occupying Cyberspace and Distributed-Denial-of-Services (DDoS) Attacks," *International Review of Law, Computers & Technology* 28, no. 1 (January 15, 2014): 98–113, doi:10.1080/13600869.2014.870638.
[20] UN Office on Drugs and Crime, "Comprehensive Study on Cybercrime" (New York, 2013), https://www.unodc.org/documents/organized-crime/cybercrime/CYBERCRIME_STUDY_210213.pdf.
[21] The Guardian, "Gary McKinnon Resource Page," *The Guardian*, 2017, https://www.theguardian.com/world/gary-mckinnon.
[22] Ahmad-Reza Sadeghi, Christian Wachsmann, and Michael Waidner, "Security and Privacy Challenges in Industrial Internet of Things," in *Proceedings of the 52nd Annual Design Automation Conference on - DAC '15* (New York, New York, USA: ACM Press, 2015), sec. 4, doi:10.1145/2744769.2747942.
[23] Leyla Bilge and Tudor Dumitras, "Before We Knew It: An Empirical Study of Zero-Day Attacks in the Real World," *Proceedings of the 2012 ACM Conference on Computer and Communications Security -- CCS'12*, 2012, 833–44, doi:10.1145/2382196.2382284.
[24] Rolf H. Weber and Evelyne Studer, "Cybersecurity in the Internet of Things: Legal Aspects," *Computer Law and Security Review* 32, no. 5 (October 1, 2016): 715–28, doi:10.1016/j.clsr.2016.07.002. - p721 and p728
[25] Shackleford, S (2013) "Toward Cyberpeace: Managing Cyberattacks Through Polcycentric Governance" American University Law Review 62(5) p1285, who lists these as 'laws and norms; market based incentives; code; self-regulation; public-private partnerships and bilateral, regional and multilateral collaboration"
[26] Ibid., 729.
[27] Urquhart and Rodden, "New Directions in Information Technology Law: Learning from Human–computer Interaction."
[28] European Commission (2013) Report on the Public Consultation on IoT Governance – p3; Weber p727; US Federal Trade Commission (2015) "The Internet of Things: Privacy and Security in a Connected World" Staff Report p7
[29] Alliance for Internet of Things Innovation WG04 (2016) "Report on Policy Issues" p34

a key global response to consider.[30] Cybercrime ordinarily entails traditional crimes enabled by IT infrastructure, like tax evasion, to true cybercrimes that would not exist but for the Internet, like bitcoin fraud, and hybrids that sit in the middle. [31] Crimes against IIoT are emerging, as are effective governance strategies. However, with criminalisation the law enforcement agencies already suffer skillset or resource deficits. These are coupled with procedural challenges of cooperating across borders to address heterogeneous, transnational cybercrimes.[32] Changes within the new EU 'Police and Justice' Data Protection Directive 2016[33] provides a framework for law enforcement agencies to cooperate and share data for investigations across borders, which may assist. Furthermore, the Council of Europe Cybercrime Convention 2001, discussed below, also contains controversial procedural powers around international cooperation and mutual assistance by states investigating and gathering evidence on crimes.[34] However, difficulties attributing attacks means criminal law may not be the most appropriate forum to redress harm. DDoS attacks, for example, could be deemed acts of cyberwar or terrorism (especially when critical infrastructure is targeted), acts of civil disobedience or protest,[35] or acts of commercial sabotage and for extortion. Adding the fear, uncertainty and doubt [36] around securing IoT to the mix, and establishing strategies that balance the benefits of IIoT with measured governance responses is tough.

In this paper, we'll consider the emergent smart energy supply chain as example of IIoT. This helps us dig into legal and, critically, technical perspectives, to reflect on security challenges posed by this trend. In part 2 we frame our analysis using the example of the smart energy supply chain, as domain where numerous new vulnerabilities may arise. We discuss relevant legal issues in these sections. In Part 3 we dig deeper into problematic elements of new laws, particularly the NIS Directive and GDPR, and in Part 4, we consider technical responses. In part 5 we offer brief conclusions.

## 2. Industrial IoT: From Exploration to Consumption

The anticipated ubiquity of networked devices embedded in infrastructure is exemplified by two current examples: smart cities and industry 4.0. The smart city movement[37] envisages urban infrastructure being upgraded to enable services like intelligent mobility[38] (e.g. congestion management, smart traffic lights, connected and autonomous vehicles) or smarter crime prevention, detection and prosecution (e.g. smart CCTV).[39] The scalability of IoT deployed in the city can frustrate effective management of security (and privacy) risks, partly due to the complexity of managing volume of data[40] and risks manifesting across

---

[30] Samantha A. Adams et al., "The Governance of Cybersecurity The Governance of Cybersecurity: A Comparative Quick Scan of Approaches in," *TILT Working Paper*, 2015, https://pure.uvt.nl/portal/files/8719741/TILT_Cybersecurity_Report_Final.pdf.
[31] David Wall, *Cybercrime : The Transformation of Crime in the Information Age* (Polity, 2007); Ross Anderson et al., "Measuring the Cost of Cybercrime: A Workshop," *Workshop on the Economics of Information Security (WEIS)*, 2012, 1–31, http://www.econinfosec.org/archive/weis2012/papers/Anderson_WEIS2012.pdf.
[32] David Wall and Matthew Williams, *Policing Cybercrime : Networked and Social Media Technologies and the Challenges for Policing*, ed. Routeledge, 2014.
[33] Directive (EU) 2016/680 of the European Parliament and of the Council of 27 April 2016 on the protection of natural persons with regard to the processing of personal data by competent authorities for the purposes of the prevention, investigation, detection or prosecution of criminal offences or the execution of criminal penalties, and on the free movement of such data, and repealing Council Framework Decision 2008/977/JHA
[34] Art 23-25
[35] Lilian Edwards, "Wikileaks, DDOS and UK Criminal Law: The Key Issues | Practical Law," *Practical Law Company*, 2010, https://content.next.westlaw.com/Document/If375d9dee81911e398db8b09b4f043e0/View/FullText.html?contextData=(sc.Default)&transitionType=Default&firstPage=true&bhcp=1.
[36] With cyberwar see Richard A. (Richard Alan) Clarke and Robert K. Knake, *Cyber War : The next Threat to National Security and What to Do about It* (Ecco, 2010).
[37] see Rob Kitchin's The Programmable City for critical engagement with the concept - http://progcity.maynoothuniversity.ie/resources/publications/
[38] Giuseppe Anastasi et al., "Urban and Social Sensing for Sustainable Mobility in Smart Cities," in *2013 Sustainable Internet and ICT for Sustainability (SustainIT)* (IEEE, 2013), 1–4, doi:10.1109/SustainIT.2013.6685198.
[39] See David Murakami Wood and Michael Carter, "Power Down," *Limn*, 2017, http://limn.it/power-down/?doing_wp_cron=1495448151.75969505310058593750000.
[40] Rolf H. Weber, "Internet of Things: Privacy Issues Revisited," *Computer Law & Security Review* 31, no. 5 (August 2015): 618–27, doi:10.1016/j.clsr.2015.07.002.

interdependent systems. As Edwards states, "smart cities are a security disaster waiting to happen".[41]

Another context causing major concern is smart manufacturing (coined as 'Industrie 4.0' in Germany or the 4th Industrial Revolution)[42]. It entails using IoT to integrate business, production and engineering processes, to enable a smarter, more flexible and responsive supply chain.[43] However, increased automation in the workplace has already been shown to pose physical risks to human co-workers when errors occur (e.g. being crushed or killed by machinery).[44] Concurrently, informational risks are prevalent, with Symantec stating manufacturing is a key target for spear phishing attacks to steal system credentials (i.e. through targeted email/communications scams), especially for industrial control systems.[45]

To establish a concrete domain to unpack possible risks and threats, we focus on a case study, the emergent *smart energy supply chain.* The new NIS Directive, enforced from May 2018, already poses challenges for the existing energy sector, like satisfying notification requirements for incidents and putting in place adequate technical and organisational compliance measures.[46] Increased networking through smart energy systems will exacerbate the risks of non-compliance, if not done with adequate foresight. Building on these concerns, we want to explore possible risks at different points in the supply chain, prioritising the following elements: drilling for raw materials on a *smart oil platform*; when transporting material from platform to land using *automated ships*; with energy generation, transmission and distribution on the *smart grid*; with *smart consumption and management* by householders. This grounds our analysis, but many of the themes discussed are translatable to other industrial IoT contexts.

## 2.1 The Digital Oilfield – IoT on Oil Platforms

Whilst data is often called the 'new oil'[47], the adoption of IoT technologies into the oil and gas industry, has been quite slow.[48] Deloitte and others cite opportunities in the emerging 'digital oilfield' like predictive maintenance driven by low cost sensors, cloud computing and big data analytics.[49] However, an awareness gap around new technologies and their applications in the industry by professionals is keeping progress slow.[50] Nevertheless, as the digital oilfield expands, forecasting risks will be necessary to ensure sustainable development in this domain (for information, safety and environmental harms).

Focusing on exploration, specifically oil platforms, we can see how IoT might be utilised in oversight of drilling operations. The goal might be sensing and analysing information about how an operation progresses to spot possible choke points (esp. those creating maintenance down time) or where components are not performing optimally (we outline an example in more detail in Part 3). Machine learning algorithms to spot trends and patterns in IoT sensor data could be deployed (similar to the different setting of mining, with Rio Tinto's autonomous

---

trucks).[51] However, the distributed, task orientated and thus heterogeneous nature of sensors means different types of data could be fed back with varying quality and at intermittent time intervals. If one firm has worked out how to cut time for a drilling operation, say, enabling them to have lower running costs and undercut their rivals at bidding stage, then this is clearly valuable to competitors. As there are many different stakeholders/competitors sharing both infrastructure (e.g. physical oil rig facilities) and components (e.g. drilling tools). This creates risks for how to maintain confidentiality in operational information that could be fed back from IoT enabled devices, guarding against advanced persistent threats (APTs) or insider threats.[52]

In response, from a practical, and security perspective, instead of aggregating IoT data into larger datasets for remote analysis, as is the current 'big data analytics' trend, the growth of industrial IoT could prompt new architectures of secure, local analysis. Not reporting raw data wholesale, but instead statistical findings, could help make IoT sensor data useful for decision making about progression and direction of operations.[53] It could also address legal compliance concerns raised by cloud based storage and appropriate safeguards being in place e.g. Privacy Shield if a US based firm, binding contract clauses, adequate third countries etc.[54] Relatedly, ensuring security mechanisms are usable for workers is important. If an IoT system is too complex to use, or the steps necessary to maintain its security have too much scope for error, then human frailties may lead to vulnerabilities. The translation from offline to online world requires traditional Computer Supported Collaborative Work (CSCW) and human factors perspectives to understand how best to design secure, usable IoT systems that workers have skills to use.[55] Furthermore, as Craggs and Rashid argues for going beyond usability towards 'security ergonomics by design' i.e. ensuring systems think about users as an integral part of the system, particularly their well-being.[56]

On the drilling platform, organisational measures to address temporal dimensions of security are important too. Ensuring secure processes are maintained over time with workers is one dimension, supported by management processes and even health and safety training. But with IoT sensors and components, there are additional risks. Securing the streams of data from IoT sensors and actuators requires maintained oversight of vulnerabilities and patching infrastructure when necessary, e.g. IoT device firmware. Preventing tampering in devices, and ensuring legacy information is not left behind moved between platforms, or even decommissioned may be necessary to ensure confidential information is not shared. Ongoing cyber-espionage activities/APTs are increasing, as high-profile campaigns like Operation Shady RAT or Operation Aurora show. These ordinarily involve targeting of state and large-scale industrial infrastructure to steal foreign intellectual property and intelligence, to assist the economic and strategic interests of the perpetrators.[57] The actors involved in these campaigns range from state sponsored hacking groups to nation states, making identification of sources, and thus appropriateness of response difficult to establish. Information from IoT on oil platforms could be another target for such campaigns, as we explore in Part 3.

### 2.1.1 Insider attacks and Unauthorised Access

Insider attacks could involve an employee accessing the rig IT system to load and execute malware or steal secrets for later sale. This could incur prosecution under unlawful access/'hacking' provisions in s1 Computer Misuse Act (CMA) 1990 (and s3 CMA for malware execution). The three part s1 CMA offence occurs when a person causes a computer to:

1) "*perform any function with intent to secure access to any program or data held in any computer*[58], *[or to enable any such access to be secured]*'";
2) where '*the access he intends to secure [or to enable to be secured] is unauthorised,*' and
3) "*he knows at the time when he causes the computer to perform the function that that is the case*'".[59]

'Securing access' means the person causes the computer to perform any function results in alteration or erasure of data, copying or moving data, causes a program to run, and so forth.[60] 'Unauthorised access'[61] is when the person is not 'entitled to control access…' and lacks consent from the one who is entitled.[62] Case law helps us unpack s1 CMA further. *Attorney General's Reference (No 1 of 1991) [1993] Q.B. 94* clarified it does not require use of a different computer for unauthorised access, but instead can be from the same computer.[63] In *DPP v Bignall*[64] authorised access to the Police National Computer was used to obtain private information. Despite the Department of Public Prosecutions (DPP) claiming this was 'unauthorised access', as their access was only meant to be for police purposes, the court found this was not a breach and "a person does not commit an offence under the 1990 Act, s1 if he accesses a computer at an authorised level for an unauthorised purpose".[65] A few years later this all changed in *R v Bow Street Magistrates Court ex Parte Allison No 2*[66] which held s1 CMA can cover activities of employees accessing data they were not authorised to.[67] The House of Lords defined scope of s1 CMA stating it "refers to the intent to secure unauthorised access to any programme or data. These plain words leave no room for any suggestion that the relevant person may say: 'Yes, I know that I was not authorised to access that data but I was authorised to access other data of the same kind.' ".[68] Insider attacks using existing login credentials would be covered by this provision, providing a route of recourse in the event of breaches.

### 2.2 Autonomous Systems in Logistics – Smart Oil Tankers

Use of autonomous systems in logistics is a clear application area for the industrial IoT. The shift towards autonomous ships (AS) is a good example, as shipbuilders across the world are investing in revolutionising transport of cargo globally (e.g. Rolls Royce).[69] Like with autonomous cars, different stages of automation will exist, and interaction between autonomous and current ships will continue.[70] For oil industry, smart oil tankers or supply vessels would be a possible application domain. Naturally, such use of AS bring a new forum for security threats to manifest. Ransomware from hackers is a big one to consider. GPS jamming, spoofing or scrambling could be used to manipulate ships or threaten to run them aground, causing financial cost and significant environmental harm (especially if the cargo is oil) unless a ransom is paid to attackers.[71] Similarly, it could present a new forum for international piracy to play out, where theft of ships is done remotely, without pirates ever needing to even set sail. Depending on the level of autonomy a ship has, insider threats would be another concern, e.g. for sabotage. On a spectrum of full to partial, manned or unmanned, this could shape to what extent insider threats manifest on board, and strict controls on who has remote access to the ship need to be maintained.

---

[58] s17(6) includes "references to any program or data held in any removable storage medium which is for the time being in the computer; and a computer is to be regarded as containing any program or data held in any such medium."
[59] CMA 1990 s1(1)(a) -(c) as Amended by Police and Justice Act 2006 c 48 Pt 5 s35 in brackets
[60] s17(2) CMA 1990
[61] s17(5) CMA 1990
[62] Amended by Criminal Justice and Public Order Act 1994 c.33 Pt XII s.162(2) (1995) section 10 relates to use of other law enforcement powers
[63] eg by "*using another person's identifier (ID) and /or password without proper authority in order to access data or a program; displaying data from a computer to a screen or printer; or even simply switching on a computer without proper authority.*" J. Zoest 'Computer Misuse Offences' (2014) *Westlaw UK Latest Update p1-*
[64] [1998] 1 Cr. App. R. 1
[65] Halsbury's Laws of England, Supplement to 11(1) (4th Ed Reissue) para 604A
[66] (AP) [2000] 2 AC 216
[67] [1999] 3 W.L.R. 620
[68] [2000] 2 A.C. 216 at 224
[69] Rolls Royce, "Autonomous Ships: The Next Steps," *Rolls Royce*, 2016, http://www.rolls-royce.com/~/media/Files/R/Rolls-Royce/documents/customers/marine/ship-intel/rr-ship-intel-aawa-8pg.pdf.

These concerns align with a wider trend towards *ransomware and extortion* campaigns, which have increased hugely in recent years,[72] leading ENISA to estimate such activity has generated a global turnover of $1bn in 2016.[73] Malware was the dominant cybersecurity threat of 2016,[74] and it has become more targeted, as financial Trojans used in the 2016 Bangladesh Bank heist show (where $81m was stolen through fraudulent transactions).[75]

The vulnerabilities posed by a recent ransomware attack, 2017's WannaCry, highlight risks of longitudinal security management in industrial IoT futures. WannaCry spread in IT systems across the globe exploiting a vulnerability in legacy system, Windows XP, which was released in 2001. The malware encrypted files stored on a system, demanding payment to decrypt and regain access. WannaCry caused widespread disruption to critical infrastructural services, for example operations and appointments at hospitals. Whilst not targeted directly at specific organisations, many services are still using XP with the vulnerability unpatched, hence it has spread far, quickly. The UK National Health Service, Spanish telecoms giant Telefonica, US logistics firm FedEx and German rail network Deutsche Bahn were all victims. For some organisations, difficulties are compounded by challenges updating systems at scale in organisations, where funding for IT services is inadequate, e.g. public sector, healthcare etc. This may be less of a risk for oil and gas sector.

In the context of smart logistics in the oil industry, utilising AS, the desire for integration between operational and management IT could increase exposure to malware. Diminishing vendor support over time, as in the case of Microsoft, would be another concern. Given the poor IoT state of emerging security practices, guarding against industrial IoT ransomware is a daunting prospect. Resources for patching vulnerabilities in a distributed network of devices, controlled by different stakeholders in a supply chain, would be logistically and practically complex. The interdependent nature of critical infrastructural systems, especially in a sector like oil with extensive outsourcing to service firms, would add another layer of difficulty. Whilst it is clear getting security right for the emerging industrial IoT is critical to ensure long term resilience and prevent substantial costs down the line, practically doing this is another matter.

---

[70] SAE international, "New SAE International Standard J3016," *SAE International*, 2016, doi:P141661.s
[71] Matt Burgess, 'When a Tanker Vanishes, All the Evidence Points to Russia' *Wired UK*, 2017 <https://www.wired.co.uk/article/black-sea-ship-hacking-russia?utm_content=bufferc8256&utm_medium=social&utm_source=facebook.com&utm_campaign=buffer> accessed 26 September 2017; Oeystein Glomsvoll and Lukasz K Bonenberg, 'GNSS Jamming Resilience for Close to Shore Navigation in the Northern Sea' (2017) 70 Journal of Navigation 33.
[72] National Cyber Security Centre and National Crime Agency, "The Cyber Threat to UK Business" (London, 2017), 5, http://www.nationalcrimeagency.gov.uk/publications/785-the-cyber-threat-to-uk-business/file. show a growth in cyber extortion; Recent Bad Rabbit Malware as another example see - http://www.bbc.co.uk/news/technology-41740768
[73] ENISA, "ENISA Threat Landscape Report 2016," 43.
[74] It is top threat in ENISA, "ENISA Threat Landscape Report 2016."; Rebecca Klahr et al., "Cyber Security Breaches Survey 2016" (London, 2016), 4, doi:10.13140/RG.2.1.4332.6324.p4 states "*the most common types of breaches experienced are viruses, spyware or malware (68%) and breaches involving impersonation of the organisation (32%)*"
[75] National Cyber Security Centre and National Crime Agency, "The Cyber Threat to UK Business," 7.

### 2.2.1 Resilience and Criminalising Material Damage

In ensuring IoT resilience, dual use tools, such as penetration test networks, have posed challenges for s3A UK CMA 1990 in the past.[76] s3A is designed to control trade in tools used for computer misuse offences by criminalising making, adaptation, supply or offer to supply articles suppliers believe it is likely to be for use/to assist in commission of CMA offences.[77] The Crown Prosecution Service have now clarified mere possession of such articles, is not an offence, without requisite intent.[78] Intent depends on factors like, normal use cases, if the article is commercially available (or only for offences) and who uses it. These wide parameters mean creating resilient IoT may still require reflection on tools used and their legal appropriateness.

Furthermore, if IIoT infrastructure is not resilient, and subject to attack, s3ZA CMA[79] criminalises causing serious damage. It applies when the accused does any unauthorised act in relation to a computer; *knowing at that time* it is unauthorised; causing, or creating a significant risk of *serious damage of a material kind*; and intending, by doing the act, to cause such damage or being reckless as to if it is caused.[80] Material damage could include to the environment or human welfare in any place or to the economy or national security of any country.[81] Material damage to human welfare is quite broad, ranging from loss of human life, illness, or injury; disruption to supply of money, food, water, energy or fuel; disruption of communications systems, transport facilities or health services.[82] When causing material damage, it is immaterial if the act causes the damage directly, or is the only or main cause of the damage.[83] Doing an act includes causing an act to be done, including if it is a series of acts. A country includes reference to a territory, and any place in, or part or region, of a country or territory.[84] This broad provision has scope for use against perpetrators of cyberattacks that cause significant damage to critical infrastructure. In the context of IIoT, if a perpetrator can be established, attacks causing black-outs impacting emergency services or the stock exchange, damage to ICS/SCADA (particularly in nuclear power stations) or general downtime for smart infrastructure, would conceivably be covered.

## 2.3 Smart Grid and Meters – Generation, Transmission/Distribution and Consumption

### 2.3.1 Generation

In the context of energy generation, the IT systems used by factories and power plants are already at risk, providing a warning for what happens when infrastructure is networked. Most concern is around vulnerabilities in industrial control systems (ICS), which come in a number of varieties but largely "*consists of a combination of control components (e.g. electrical, mechanical, hydraulic, pneumatic) that act together to achieve an industrial objective (e.g. manufacturing, transportation of matter or energy)*".[85] Traditionally, ICS are 'air-gapped' (i.e. not connected to the internet) to limit vulnerability to external attacks which can cause physical harm. This means, threats often emerge from actors physically booting vulnerabilities locally, e.g. via USB. But with the growth of the industrial IoT and networked integration across systems, this safeguard is being removed.[86] Traditional phishing campaigns are a risk,[87] and an attack on Ukrainian electricity distribution companies Prykarpattya Oblenergo and Kyiv Oblenergo led to blackouts and power outages and affected over 220,000 customers and utilised malware distributed through phishing emails and malicious Microsoft Word files.[88] Malware linked to these Ukranian attacks, Industroyer, is particularly dangerous because it enables control of substation switches and circuit breakers.[89]

---

[76] as amended by the s41 Serious Crime Act 2015; See Edwards (2010) at section 'is merely downloading the LOIC a crime?'
[77] s3A (1-3) CMA – other offences of s1, s3 or s3ZA CMA 1990
[78] Crown Prosecution Service Legal Advice on Computer Misuse Act 1990 available at http://www.cps.gov.uk/legal/a_to_c/computer_misuse_act_1990/
[79] Added by s41 Serious Crime Act 2015
[80] s3ZA (1)
[81] s3ZA (2)
[82] s3ZA (3) CMA 1990
[83] s40(4)
[84] s40(5)
[85] Keith Stouffer et al., "Guide to Industrial Control Systems (ICS) Security," *NIST*, 2015, doi:10.6028/NIST.SP.800-82r2.2-1

Exploitation of zero day vulnerabilities against ICS used in power plants and factories like SCADA,[90] a type of ICS defined by US Standards agency NIST as "*systems [that] are used to control dispersed assets where centralized data acquisition is as important as control…*" are also prevalent.[91] A recent SCADA hack targeting a German steel mill that suffered physical damage in 2015.[92] However, arguably the highest profile targeted ICS attack was the state sponsored 2010 Stuxnet worm attack (allegedly from the US and Israel)[93] on the Iranian Natanz nuclear enrichment plant. It targeted a specific Siemens ICS, using a combination of fake authentication certificates and zero day exploits[94] to reach its target and deploy a complex payload designed to vary speed at which uranium enrichment centrifuges spin, thus destroying them. The payload slowed production at the plant, as centrifuges had to be replaced more quickly. Ultimately, it aimed to delay production of purportedly nuclear weapons using enriched uranium as part of the Iranian Nuclear program.[95] For industrial IoT, these cases highlight the need for careful consideration about what should and should not be networked and connected to the Internet, relative to costs and benefits (both economic and security).

### 2.3.1.1 Addressing ICS Hacks

Targeting critical civilian infrastructure, like ICS, as the 'battlefield' for playing out international tensions complicates navigation of this domain.[96] The international law on cyberwarfare may come to the fore, both with the jus ad bellum and jus in bello.[97] The NATO Cooperative Cyber Defence Centre of Excellence in Tallinn has sought to create clarity, through the Tallinn Manuals. These interpret application of public international law to cyber operations during armed conflict[98] and more recently, during peacetime.[99] They focus on use of force and self-defence in Article 2(4) and Article 51 of the UN Charter, beyond the original scope of armed attacks causing kinetic damage. As mentioned before, attributing attacks is tricky, especially when it's the basis for justifying action between nation states. Network traffic can be masked and routed via several countries to hide the identity of perpetrators, making establishment of state responsibility for cyber-attacks difficult.[100] Furthermore, given the messy crossover between cyber- war, crime, espionage, and terrorism, to name a few, holding nation states responsible for acts of groups acting autonomously within their borders is tough. This is especially if such groups do so without knowledge or authority of the armed forces. Questions of proportionality of responses to interstate cyber-attacks also requires political and ethical reflection. Even if kinetic attacks in response to cyber-attacks can be deemed legal,[101] is it morally correct to do so? With states designing and building cyber weapons like Stuxnet, is a cyber arms trade treaty needed to control weapon use or even for a ban banning some, as with nuclear weapons or chemical weapons?[102] Nevertheless, despite all these difficult questions, some experts suggest the risk are overstated.[103] Instead, perhaps we need to refocus on the more mundane threats to power grids - electrocuted squirrels and birds causing outages.[104]

---

[86] ENISA, "Protecting Industrial Control Systems: Recommendations for Europe and Member States" (Heraklion, 2011); Barak Perelman, "Air Gap or Not, Why ICS/SCADA Networks Are at Risk | SecurityWeek.Com," *Security Week*, 2016, http://www.securityweek.com/air-gap-or-not-why-icsscada-networks-are-risk.
[87] ENISA Smart Grid Recommendations 2012
[88] HM Government, "National Cyber Security Strategy 2016-2021," 2016, https://www.gov.uk/government/uploads/system/uploads/attachment_data/file/567242/national_cyber_security_strategy_2016.pdf.
[89] John Leyden, "Move Over, Stuxnet: Industroyer Malware Linked to Kiev Blackouts • The Register," *The Register*, 2017, https://www.theregister.co.uk/2017/06/12/industroyer_malware/.
[90] Vinay M. Igure, Sean A. Laughter, and Ronald D. Williams, "Security Issues in SCADA Networks," *Computers and Security* 25, no. 7 (2006): 498–506, doi:10.1016/j.cose.2006.03.001.
[91] Stouffer et al., "Guide to Industrial Control Systems (ICS) Security." s2-5 continues "[examples] *distribution systems such as water distribution and wastewater collection systems, oil and natural gas pipelines, electrical utility transmission and distribution systems, and rail and other public transportation systems…SCADA systems are designed to collect field information, transfer it to a central computer facility, and display the information to the operator graphically or textually, thereby allowing the operator to monitor or control an entire system from a central location in near real time.*"
[92] Kim Zetter, "Car Wash Hack Can Strike Vehicle, Trap Passengers, Douse Them With Water - Motherboard," *Motherboard*, 2017, https://motherboard.vice.com/en_us/article/bjxe33/car-wash-hack-can-smash-vehicle-trap-passengers-douse-them-with-water.z
[93] Ellen Nakashima and Joby Warrick, "Stuxnet Was Work of U.S. and Israeli Experts, Officials Say - The Washington Post," *The Washington Post*, 2012, https://www.washingtonpost.com/world/national-security/stuxnet-was-work-of-us-and-israeli-experts-officials-say/2012/06/01/gJQAlnEy6U_story.html?utm_term=.9ee2a60c2170.
[94] i.e. unpatched vulnerabilities in IT systems that can be exploited. A market exists in buying these exploits before they are patched by vendors
[95] Kim Zetter, "How Digital Detectives Deciphered Stuxnet, the Most Menacing Malware in History," *Wired*, 2011.
[96] For even more detail the Special Edition on Cyberwarfare of *Journal of Conflict and Security Law* (2012) - Vol 17:2 - http://jcsl.oxfordjournals.org/content/17/2.toc
[97] H H Dinniss, *Cyber Warfare and the Laws of War*, Cyber Warfare and the Laws of War (Cambridge: Cambridge University Press, 2012), doi:10.1017/CBO9780511894527.
[98] Split into 2 parts – Part I International Cybersecurity Law (i.e. primarily the jus ad bellum) with state attribution (Rules 6-9); Use of Force (10-12); Self Defence (13-17); then Part II on Law of Cyber Armed Conflict (i.e. primarily the jus in bello) with detailed rules on cyber weapons, legitimate targets, cyber espionage and the nature of attacks (Rules 25-66)
[99] CCD COE NATO, *Tallinn Manual 2.0 on the International Law Applicable to Cyber Operations*, 2nd ed. (Tallinn: Cambridge University Press, 2017); CCD COE NATO, *Tallinn Manual on the International Law Applicable to Cyber Warfare* (Tallinn: Cambridge University Press, 2013).

### 2.3.2. Transmission

Industry and government are driving the shift towards smart grids, i.e. *"an upgraded energy network to which two-way digital communication between the supplier and consumer, smart metering and monitoring and control systems have been added"*.[105] The grid aims to create more efficient energy production by industry, smarter consumption by citizens and works towards domestic, regional and international $CO_2$ emission reduction targets.[106] Levelling off inefficient peaks in consumer demand is a goal, relying on more than just understanding consumer behaviours, but changing them. Pricing is one mechanism, and consumers could be incentivised to change habits through time of use (TOU) tariffs i.e. where electricity pricing changes at different times of the day (Goater, 2014, p3).[107] There are security risks here in malicious manipulation of supply and demand, for economic loss by both providers and consumers (especially with consumption being measured by smart meters, more on this below). Smart grid security has had strategic attention from bodies like ENISA, with numerous best practice documents requiring resilience by design technical means (e.g. end to end security) and organisational ones (e.g. risk assessments).[108] A particular risk that is hard to manage is distributed denial of service (DDoS) attacks, often facilitated by botnets.

There is an extensive number of botnets and a 2013 UN Office on Drugs and Crime Comprehensive Study on Cybercrime estimated around 1 million botnet C&Cs globally, with high volume clusters in Eastern Europe, Central America, and the Caribbean.[109] Devices compromised by malware become infected "zombie" units, enslaved to a command and control server which remotely control their behaviour on demand. Botnets are then put to work, often for hire, for DDoS campaigns, by a range of actors, from organised crime groups to script

---

[100] Jeffrey Carr, "Responsible Attribution: A Prerequisite for Accountability," *The Tallinn Papers: A NATO CCD COE Publication on Strategic Cyber Security*, 2014, https://ccdcoe.org/sites/default/files/multimedia/pdf/Tallinn Paper No 6 Carr.pdf.
[101] David Alexander, "U.S. Reserves Right to Meet Cyber Attack with Force | Reuters," *Reuters*, 2011, http://www.reuters.com/article/us-usa-defense-cybersecurity-idUSTRE7AF02Y20111116.
[102] Arimatsu (2012) "A Treaty for Governing Cyber-Weapons" *CCD COE Cycon* http://www.ccdcoe.org/publications/2012proceedings/2_3_Arimatsu_ATreatyForGoverningCyber-Weapons.pdf
[103] Rid, *Cyber War Will Not Take Place*.
[104] Cleve Wootson Jr, "Most Cybersecurity Experts Are Worried about Russian Hackers. One Says: Look, a Squirrel!," *The Washington Post*, January 2016. http://www.cybersquirrel1.com/#;
[105] Communication 2012/148/EC Section 3(a) Definitions
[106] The UK DECC need to reduce CO2 by 80% by 2050 from 1990 levels, in line with the UK Climate Change Act 2008 and Energy Act 2011
[107] e.g. cooking between 5 and 8pm or having showers between 6-8am
[108] ENISA, "Smart Grid Security Recommendations" (Heraklion, 2015), http://www.enisa.europa.eu/activities/Resilience-and-CIIP/critical-infrastructure-and-services/smart-grids-and-smart-metering/ENISA-smart-grid-security-recommendations.
[109] UN Office on Drugs and Crime 'Comprehensive Study on Cybercrime' (Vienna: UNODC 2013) p33

kiddies.[110] DDoS attacks flood servers with requests, meaning services hosted on targeted servers are knocked offline temporarily, but DDoS attacks are not permanent and impacts often resolved once servers are brought back online.[111] For the smart energy grid, DDoS attacks could impact transmission and distribution networks, leading to power outages and associated black outs, where physical safety is at stake.[112] Furthermore, it can impact flows of information between consumers and producers, where costs go beyond downtime but also disrupting production schedules, leading to significant economic, safety or political costs as second order effects are felt down the supply chain.

### 2.3.2.1. Confronting DDoS

Whilst Internet Service Providers have a role to play in monitoring and throttling high volumes of traffic, criminally tackling DDoS pushes us to s3 CMA 1990. It covers '*unauthorised acts with intent to impair, or with recklessness as to impairing, the operation of a computer…*'. Such acts (or a series of acts) can involve temporary impairment, prevention or hindering operation of a computer, being indiscriminate towards computers, programmes or data. As DDoS attacks do not ordinarily cause permanent damage to the server, merely knocking it offline temporarily, they still come within the scope of s3. The *DPP v Lennon* (2006) examined[113] a mail bombing campaign committed by Lennon against a former employer's servers.[114] The court accepted sending emails was a modification to a computer (before 2006, s3 required unauthorised 'modification' instead of an 'act').[115] The case focused on the authority for this act, especially when sending emails is ordinarily an authorised activity. The court held that the implied consent of a user to receive emails is not without limits,[116] and such consent does not stretch to cover situations where the purpose of emails is to overwhelm the system, as is the case with DDoS too. Lord Justice Keene stated the recipient "*does not consent to receiving emails sent in a quantity and at a speed which are likely to overwhelm the server. Such consent is not to be implied from the fact that the server has an open as opposed to a restricted configuration.*"[117] Accordingly, there is precedent around DDoS type attacks flooding a server with requests, and this would criminalise DDoS attacks against IIoT infrastructure e.g. targeting components of the supply chain.

### 2.3.3. Consumption

As part of the smart grid, homes around Europe (and the world) are being fitted with smart meters i.e. '*electronic systems that can measure energy, consumption, providing more information than a conventional meter, and can transmit and receive data using a form of electronic communication*'.[118] In the UK, the Smart Meter Implementation Programme (SMIP) here run by the (former) Department of Energy and Climate Change (DECC), now BEIS,[119] with an installation target of 53 million gas and electricity smart meters across the UK by 2020. It is part of the wider EU shifts, namely the EU Third Energy Package[120] and specifically, Directive 2009/72/EC which requires 80% of Europe to be using smart meters by 2020. This means around 200 million electricity smart meters (72% of all European consumers) and 45 million gas meters.[121] SMIP has been delayed extensively with issues around cost, impacts on vulnerable populations and lacking transparency, to name a few.[122] Nevertheless, by 31 March 2016, official UK statistics show there are 2.75 million smart meters across UK operating in smart mode, representing 5.8% of total domestic meters in UK (DECC, 2016).[123]

---

[110] Giles Hogben (ed) *Botnets: Detection, Measurement, Disinfection and Defence* (Heraklion: ENISA 2013)
[111] See legal dimensions in Lilian Edwards "Dawn of the Death of Distributed Denial of Service: How to Kill Zombies",2006, *Cardozo Arts and Entertainment Law Journal* 24(1), 23-59
[112] Andy Greenberg, 'Summer of Discontent: Dragonfly 2.0 Hacking Campaign Targeted US and European Power Grids', *Wired*, 2017, https://www.wired.co.uk/article/hackers-power-grids-uk-symantec.
[113] Using Avalanche v3.6 program
[114] The emails were made to appear to come from a manager within the company
[115] Amended by Police and Justice Act 2006 s36
[116] See s17(8)(b) CMA on definition of an 'unauthorised act'
[117] *DPP v Lennon* [2006] EWHC 1201 (Admin) at 14
[118] from Article 2 Energy Efficiency Directive (2012/27/EU)
[119] UK Department for Business Energy and Industrial Strategy
[120] Electricity Directive (2009/72/EC) Annex I.2

At the consumer level, threats stem from smart meters and home energy management tools becoming compromised and exploited. Poorly secured IoT devices often use default passwords and thus have scope for data breaches as they interface with other IoT devices. This can lead to individual privacy harms, for example by compromised data directly or indirectly making patterns of daily life and occupancy visible to external actors. Smart thermostats and in-home displays to energy efficient smart lighting and washing machines share domestic networking, thus each can bring risks into the home. Another near future concern is security vulnerabilities in agent based systems deployed in the smart grid to assist with demand side management, e.g. with dynamic price tariffing. In the future, to level peak demands on the smart grid, prices may be changed rapidly to encourage consumption at different times. Due to complexity of managing this, consumers may need software agents negotiating tariffs on their behalf.[124] Compromised agents could create substantial energy bills for consumers, and again, be another forum for ransom and extortion, e.g. pay us £500 or pay a £750 energy bill.

The compromised IoT infrastructure, much like more traditional 'zombie PCs', can be implicated in botnets, particularly unsecured consumer grade systems. The Shodan search engine shows unsecured IP connected devices, like baby cams,[125] and the UK NCA argues, *"the Shodan search engine reveals, for example, over 41,000 units of one insecure model of DVR are connected to the Internet as of January 2017"*.[126] These are being exploited, and recent DDoS attacks on a domain name service (DNS) company were mediated, in part, by the Mirai IoT botnet made up of compromised IP connected security cameras and digital video recorders (DVRs).[127] In 2017, more IoT botnets wer found, including one targeting IP Cameras specifically, Persirai,[128] an IoT worm Hajime[129] and the Reaper botnet, created by actively hacking software instead of just hunting for default passwords.[130]

### 2.3.3.1 Tackling Botnets

In fighting botnets, strategy argued by ENISA is to prevent new infections, break up existing botnets and minimise financial gains made from them.[131] These new IoT botnets are covered by the Council of Europe's longstanding Cybercrime Convention 2001 (CCC '01). IoT devices are computer systems within CCC '01's definition i.e. 'any device or group of interconnected or related devices, on or more of which, pursuant to a program, performs automatic processing

---

[121] European Commission, "Benchmarking Smart Metering Deployment in the EU 27 with a Focus on Electricity" (Brussels, 2014).
[122] Public Accounts Committee, "Twelfth Report: Update on Preparations for Smart Metering" (London, 2014).
[123] Meters operated by big energy firms eg British Gas, SSE, E.On etc
[124] Tom A. Rodden et al., "At Home with the Agents," in *Proceedings of the SIGCHI Conference on Human Factors in Computing Systems - CHI '13* (New York, New York, USA: ACM Press, 2013), 1173, doi:10.1145/2470654.2466152.
[125] JM Porip, 'How to Search the Internet of Things for Photos of Sleeping Babies', *Ars Technica*, 2016, https://arstechnica.co.uk/security/2016/01/how-to-search-the-internet-of-things-for-photos-of-sleeping-babies/ .
[126] National Cyber Security Centre and National Crime Agency, "The Cyber Threat to UK Business."
[127] Ibid.
[128] John Leyden, "Another IoT Botnet Has Been Found Feasting on Vulnerable IP Cameras • The Register," *The Register*, 2017, https://www.theregister.co.uk/2017/05/10/persirai_iot_botnet/.
[129] Waylon Grange, "Hajime Worm Battles Mirai for Control of the Internet of Things," *Symantec*, 2017, https://www.symantec.com/connect/blogs/hajime-worm-battles-mirai-control-internet-things.
[130] Greenberg A (2017) The Reaper IoT Botnet Has Already Infected A Million Networks, Wired https://www.wired.com/story/reaper-iot-botnet-infected-million-networks/
[131] Jan Gassen, Elmar Gerhards-Padilla, and Peter Martini, 'Botnets: How to Fight the Ever-Growing Threat on a Technical Level' in Heli Tirmaa-Klaar et al., Botnets (Springer 2013). p34

of data'.[132] By way of background, it seeks to create "*a common criminal policy aimed at the protection of society against cybercrime, inter alia by adopting appropriate legislation and fostering international co-operation*".[133] It looks for harmonisation by signatories providing domestic legislation on five offences, including hacking, computer based fraud or distributing illegal content.[134] As of March 2017, it has 52 overall ratifications. where the, ratified in 2011 and the Convention came into force in September 2011.[135] The UK signed in 2001, ratified in 2011 and covers requirements through amendments to the Computer Misuse Act 1990.

The relevance of CCC '01 has been questioned, primarily due to aging definitions and classifications of offences not encapsulating current attacks (like ransomware)[136]. In keeping it up to date, the Cybercrime Convention Committee (T-CY) has issued guidance notes[137] and they state botnets fall within CCC '01 remit because "*computers in botnets are used without consent and are used for criminal purposes and to cause major impact*".[138] Accordingly, they are covered by many provisions of CCC '01, such as Article 2 on illegal access (due to the malware creating the zombie for the botnet) and Article 4 on data interference (as it alters data and sometimes delete, damage, deteriorate or suppresses it).[139] Information sharing and computer early response teams (CERTS) have an important role to play tackling botnets. We discuss CERTs further below, but the UK CERT and CiSP[140] information sharing scheme have made progress fighting bots.[141]

### 3. New Legal Requirements

Against the technical threat backdrop, we also have a range of regulatory considerations to consider. Organisations providing critical infrastructure have an increasing role in addressing cybersecurity risks. A key challenge is balancing these legal obligations with the commercial drive towards the industrial IoT. The EU Network and Information Security (NIS) Directive 2016,[142] enforced from May 2018,[143] defines obligations by establishing minimum pan-EU harmonised standards. EU member states need to adopt national measures and implementation strategies, particularly for cross-border cooperation. A network of computer security incident response teams (CERTS) and a strategic cooperation group to bring states together to share information about attacks are two examples. Short term, the UK remains committed to the NIS Directive, but long term, the nature of future cooperation remains unsettled.[144]

### 3.1 NIS Directive 2016: Security of Essential Services

With NIS, EU member states need to identify the operators of 'essential services' in their territory, from across energy, transport, banking, financial markets and health sectors.[145] This includes bodies such as energy operators involved in supply, distribution and storage of natural resources (e.g. oil pipelines, refineries and rigs); transportation providers (e.g. air carriers, intelligent transport systems or traffic management); banking (e.g. credit institutions); financial trading (e.g. stock markets); and healthcare providers (e.g. hospitals or clinics).

---

[132] Article 1 Cybercrime Convention
[133] See Preamble of Cybercrime Convention
[134] Chapter II Section 1
[135] http://www.coe.int/en/web/conventions/full-list/-/conventions/treaty/185/signatures?p_auth=VOztoKSJ
[136] Weber and Studer, "Cybersecurity in the Internet of Things: Legal Aspects."
[137] Committee on Cybercrime Convention, 9th Plenary of the T-CY (2013) Guidance Notes 2-7
[138] Committee on Cybercrime Convention, 8th Plenary (2013) https://rm.coe.int/16802e7132 p6
[139] Committee on Cybercrime Convention 8th Plenary (2013) p7
[140] UK Cyber Security Information Sharing Partnership - https://www.ncsc.gov.uk/cisp
[141] Samantha A. Adams et al., "The Governance of Cybersecurity The Governance of Cybersecurity: A Comparative Quick Scan of Approaches in," 58–60.
[142] NIS Directive EU 2016/1148 (NIS)
[143] NIS Article 25
[144] http://www.out-law.com/en/articles/2017/january/network-and-information-security-directive-will-be-implemented-in-the-uk-despite-brexit-vote-government-confirms/
[145] NIS Annex II

Article 14 NIS states operators of essential services need to put in place appropriate, proportionate technical and organisational measures to address risks posed to systems, relative to the state of the art. They need to take measures to ensure continuity of service and prevent/minimise impacts of incidents. They also need to notify relevant authorities (e.g. a regulator or computer emergency response team),[146] without undue delay, about incidents that affect their ability to provide their services, including cross border dimensions. Number of users affected by disruption of the service, duration of incident and geographical spread of area affected by the incident should be considered. This information may be shared with other member states so they can respond too.

Curiously, it also extends to three specific digital services, online marketplaces, search engines, and most interestingly here, cloud computing services.[147] With the latter, similar provisions to Article 14 on technical and organisational measures exist in Article 16 NIS, but add that the following factors should also be taken into account: (a) the security of systems and facilities; (b) incident handling; (c) business continuity management; (d) monitoring, auditing and testing; (e) compliance with international standards. For determining if an incident is substantial, duration and geographical spread remain. However, impact on economic and societal activities, extent of disruption, and number of users relying on their services to provide their own services also need to be reflected. With digital services, the public may be notified where necessary by authorities. Article 16 does not apply to micro and small businesses.[148] With both Article 14 and 16, member states need to make sure that there are appropriate regulatory powers (including setting penalties)[149] for authorities to enforce the rules.[150]

In criticising NIS, Weber argues are the nature of appropriate and proportionate technical and organisational measures (APTO) measures remains nebulous; the exemption for SMEs, hardware and software providers is too much, as it excludes many important actors from the law; and, given reputational harms associated with reporting breaches, implementation of mandatory breach notification requirements may be lacklustre (Weber, p726). We provide exploration of APTOs

### 3.2 Computer Emergency Response Teams[151]: Managing IIoT Vulnerabilities

Any growth of industrial IoT in critical infrastructure, needs to ensure it complies with these substantive requirements in NIS around risk mitigation and notification requirements. In the context of distributed IoT devices, this could be a tall order. At a strategic level, alongside NIS, both the UK/EU Cybersecurity Strategies[152] cite the importance of CERTs in quickly addressing cybersecurity risks. Hence, at a societal level, in conjunction with ENISA, CERTs have a key role in training exercises, issuing guidance, and ensuring cooperation across borders for industrial IoT. Raising awareness and finding strategies to address nascent security risks will be a key role in the future.

Patching industrial IoT vulnerabilities is likely to be a huge undertaking, even if resources and planning are invested. The UK Cybersecurity Strategy argues that vulnerabilities are growing due to the number of systems going online, creating more threat vectors but poor cyber hygiene practices by the population, such as not using antivirus software, the lack of security skills across society, from the general-public to public and private sectors and also the continued use of unpatched legacy IT systems are primary concern.[153] The UK National Crime Agency echo the latter point, concerned that despite widespread publicity of many vulnerabilities, like Heartbleed, they have not been fully patched and remain.[154] This enables nation states to take advantage of the old vulnerabilities, utilising less sophisticated approaches to leverage hacks to steal intellectual property or state secrets, and leaving more sophisticated tools for when truly necessary.[155] How these vulnerabilities manifest in industrial IoT contexts remains to be seen.

---

[146] Called computer security incident response teams (CSIRTS) in NIS
[147] NIS Annex III
[148] Art 16(11) NIS Directive
[149] NIS Article 21
[150] NIS Articles 15 and 17
[151] See NIS Article 9 for more on CSIRTS
[152] Ian Levy, Active Cyber Defence – Tackling Cyber Attacks in the UK, NCSC, 2016 https://www.ncsc.gov.uk/blog-post/active-cyber-defence-tackling-cyber-attacks-uk

### 3.3 EU General Data Protection Regulation 2016: Notification Requirements and Workers' Personal Data

The new EU General Data Protection Regulation 2016 (GDPR) also needs to be considered here as it includes provisions on security of personal data.[156] It includes new notification rules around personal data breaches i.e. '*a breach of security leading to the accidental or unlawful destruction, loss, alteration, unauthorised disclosure of, or access to, personal data transmitted, stored or otherwise processed.*"[157] Any data controllers who suffers a personal data breach needs to notify the UK data protection regulator, the UK ICO, within 72 hours of discovery of the attack.[158] They need to provide quite detailed information in a very short period of time, including:

a) "the nature of the personal data breach including where possible, the categories and approximate number of data subjects concerned and the categories and approximate number of personal data records concerned.
b) communicate the name and contact details of the data protection officer or other contact point where more information can be obtained;
c) describe the likely consequences of the personal data breach;
d) describe the measures taken or proposed to be taken by the controller to address the personal data breach, including, where appropriate, measures to mitigate its possible adverse effect."[159]

In addition, they need to let the data subject know about the breach too, in a clear and plain manner, without undue delay (but not within 72 hours) is likely to pose high risks to their rights and freedoms.[160] However, they do not need to do this, if the following three conditions are met:

(a) "the controller has implemented appropriate technical and organisational protection measures, and those measures were applied to the personal data affected by the personal data breach, in particular those that render the personal data unintelligible to any person who is not authorised to access it, such as encryption;

---

[153] Government, "National Cyber Security Strategy 2016-2021."
[154] National Cyber Security Centre and National Crime Agency, "The Cyber Threat to UK Business," 9.
[155] Ibid., 7.
[156] GDPR Article 32
[157] GDPR Article 4(12)
[158] GDPR Article 33
[159] GDPR Article 33 (3)
[160] GDPR Article 34

(b) the controller has taken subsequent measures which ensure that the high risk to the rights and freedoms of data subjects referred to in paragraph 1 is no longer likely to materialise;

(c) it would involve disproportionate effort. In such a case, there shall instead be a public communication or similar measure whereby the data subjects are informed in an equally effective manner."[161]

Given the differentiated notification provisions here, end users are often likely to be finding out about data breaches through news stories or public messages from companies more often,[162] as data breaches in 2016 were 45% higher than in 2014.[163] As discussed above, the insecurity in the domestic IoT interfaces with industrial IoT, insofar as it becomes part of botnets that are then used to attack critical infrastructure. Cloud service providers have extensive security obligations under NIS, and the design of many IoT systems is orientated towards sensing data then aggregating it in the cloud for analytics to provide contextually relevant service. So, when IoT products utilise cloud services when handling personal data, both NIS and GDPR obligations could come to the fore. In terms of putting in place NIS mandated organisational and technical measures to ensure security, coupled with notification obligations from GDPR, the case gets stronger for IoT systems to be designed in a manner where cloud storage is not the dominant approach. As mentioned above, local data storage and analytics would help organisations avoid a lot of these difficult compliance requirements and enable more controlled and sustainable security architecture too. Technologies like personal information management systems[164] are useful for protecting consumers' personal data, but they also have much to offer for industrial IoT too, in terms of providing confidentiality or limiting access to sensitive information. Hardware level trusted execution environments (i.e. a secure space on the chip) can also play a role in industrial IoT, attesting to identities of devices in widely distributed systems.[165]

Another relevant provision of GDPR for IIoT is Article 32 on security of processing. IIoT is primarily about integrating and tracking information at different points in goods or service supply chains, but workers are also a key part of this process. IIoT can disrupt their current work practices by introducing greater oversight by observing how they complete tasks, spotting inefficiencies and trying to increase productivity through automation, where possible. Worker personal data is manifest in the mix meaning information privacy obligations still need to be considered and how IIoT systems impact their rights.[166] In some jurisdictions, a combination of labour laws, unionisation and system design could tackle negative impacts of IIoT and automation, as occurred with the Scandinavian School of Participatory Design movement when IT was introduced into workplaces in the 1970s.[167] However, we focus here on a specific security provision in Article 32 GDPR that deals with 'security of processing'. Personal data of workers needs to be handled in a secure manner, but the shortcomings in IIoT security may see them implicated in data breaches and other privacy harms. To prevent this, IIoT deployments need to take stock of Article 32 GDPR. Broadly, it states appropriate 'technical

---

and organisational measures' need to be taken by controllers and processors to protect rights and freedoms of data subjects. This has to take into consideration: the 'state of the art', 'costs of implementation', 'nature, scope, context and purposes of processing' and 'likelihood and severity' of risk to their rights. They give examples such as pseudonymising or encrypting data, testing the resilience, integrity, confidentiality and availability of processing systems, or restoring access and availability of data quickly after an incident. If they abide by codes of practice, that can be a demonstration of compliance. Article 32 again surfaces the need for technical and organisational measures, which is a turn to the technical community to play a critical role in regulating the risks. Alongside technical requirements of creating functioning IIoT systems, such legal requirements increasingly need to be thought about in early stages of the system design process.[168] However, like with privacy by design, this is much easier said than done, and extensive work will be needed to both situate the role of developers and security engineers in regulation. Work is needed to support their efforts to embed compliance mechanisms in design, through translation of law into technically actionable measures or through new tools to better surface their obligations.[169]

**4. Engineering the Industrial IoT: Appropriate Technical and Organisational Measures**
As we've seen above, law is increasingly focusing on the role of technical and organisational measures to address cybersecurity risks. This is both in NIS, for critical infrastructure providers or GDPR, for security of personal data. Accordingly, the law is bringing technical professionals to the fore, and there is a growing space for technical responses for IIoT security, to supplant legal approaches. We've already hinted towards the importance of distributed storage approaches above and a growing need for edge computing too.[170] The bandwidth and networking challenges of sending large volumes of data (e.g. from autonomous vehicles) from sensors to the cloud for central analytics mean conducting analytics locally and sending back results is increasingly attractive.[171]

In reflecting on these issues and considering routes redressing the risks stemming from IoT, established practices in IT architecture design could be considered. Examples could include:

- Keep data distributed, as opposed to centralising the data into one, more vulnerable central storage point.
- Keeping data encrypted both when stored and when being sent over networks.
- Keeping controls on access by third parties through white lists and credentialing.
- Using local storage and analytics, where the raw data does not leave the hardware, and any analytics can be run locally (with results relayed back to relevant stakeholders).

Returning to our example of Industrial IoT on oil platforms in Part 2, we now explore putting networked sensors into devices in more detail. On the platform, monitoring integrity of components like valves on blowout preventers, connectors on hoses or structure of derrick frames would be important to save on possible down time by spotting issues early and observing performance to schedule servicing or replacement. Accordingly, sensors may be installed to:

---

[168] Urquhart and Rodden, "New Directions in Information Technology Law: Learning from Human–computer Interaction"; M Dennedy, J Fox, and T Finneran, *Privacy Engineer's Manifesto*, JOUR (New York: Apress, 2014); Irit Hadar et al., "Privacy by Designers: Software Developers' Privacy Mindset," *Empirical Software Engineering*, April 30, 2017, 1–31, doi:10.1007/s10664-017-9517-1.
[169] George Danezis et al., "Privacy and Data Protection by Design - from Policy to Engineering," *European Network and Information Security Agency* (Heraklion, 2014); Ewa Luger et al., "Playing the Legal Card: Using Ideation Cards to Raise Data Protection Issues within the Design Process," JOUR, in *Proceedings of the ACM CHI'15 Conference on Human Factors in Computing Systems*, vol. 1, (2015), 457–66, doi:10.1145/2702123.2702142.
[170] Carmela Troncoso et al., "Systematizing Decentralization and Privacy: Lessons from 15 Years of Research and Deployments," *Proceedings on Privacy Enhancing Technologies* 2017, no. 4 (2017): 307–29, doi:10.1515/popets-2017-0052.
[171] Mortier et al., "Personal Data Management with the Databox : What ' S Inside the Box ?"; Wenting Zheng et al., "Opaque: An Oblivious and Encrypted Distributed Analytics Platform," accessed November 30, 2017, https://people.eecs.berkeley.edu/~wzheng/opaque.pdf.

- Monitor sudden changes in temperature;
- Pipe pressure;
- Oil flow speed;
- Fatigue in components;
- Strength of joints in pipelines;
- Analysis of chemical composition of quality of oil etc.

Depending how these sensors are networked, and how vulnerable they are to attacks, this shift could create new threat vectors. Taking a few examples below, we pose a range of questions to consider:

**Networking Approaches** –Existing network infrastructure on rigs for getting data back onshore, will be important, or at least from installed sensor devices to the rig. What costs might be associated with telecoms provision to transfer data in remote locations like the North Sea or Siberia? Will the system use networking approach will be used (e.g. star, mesh)? How secure will these be? What protocols will be used for networking? Ensuring encryption during transmission will be key, how much bandwidth is available for relaying information will dictate the granularity of data that can be sent? and how regularly?

**On-board Storage Capacity** – how often do the sensors need to be 'emptied', with associated costs for servicing by staff (e.g. divers if they are remotely on the seabed? Distributed nature of the IoT system could be beneficial from a security perspective, but only if done properly.

**Computational Capabilities** – design decisions about processing power on devices dictate scope for local analytics vs the need to send to the 'cloud' for analysis on servers with greater computational capacity. Power and battery life of sensors could be a problem too, as adding processors would drain power more quickly. These decisions could create new threat vectors, for example around cloud security for confidential data.

**Resilience of Devices** – temporal considerations are key, as the harshness of the environment may impact physical security of devices and sensors. In terms of software, ability to update and patch vulnerabilities in the code may be difficult too, if devices are hard to access e.g. deep underwater.

**5. Conclusions and Further Work**

As we have explored, the emergence of CBS, as encapsulated by the industrial IoT, can bring many new security vulnerabilities. The context of smart energy infrastructure, from resource exploration to energy consumption, helped us unpack some of the key challenges. Engineering dimensions around sensors are useful to reflect when analysing how regulatory frameworks might shape the nature of the industrial IoT. From a legal perspective, balancing the obligations in NIS with the desire for industrial IoT is one challenge, the need for guidance from CERTS and authorities on IoT is another. Ultimately, security requirements from NIS and GDPR around cloud may also prompt growth of alternative architectures for the industrial IoT. How these may manifest legally, commercially and technically is an area in need of further research. In going forward, there needs to be an increased focus on understanding the implications of the shift of infrastructure from offline to online; how to handle temporal dimensions of security; how best to address implementation gaps for best practice; and how to engage with the infrastructural complexity of critical systems. To conclude, we consider each in more detail in turn.

**From Offline to Online** - New risks and vulnerabilities arise from networking infrastructure that is traditionally 'offline' being put online, and automating industrial processes that may traditionally have greater human oversight. Current best practices may not translate when automated, as security implications of putting something online that was not formerly networked might not be fully anticipated.

**Temporality and Security** – Planning and building security into goods and services requires motivation, oversight and forecasting of risk. Managing security over time is a complex variable to consider. The distributed nature of IoT being integrated at scale into critical, industrial infrastructure poses questions about effective longitudinal strategies. Ensuring data security considerations are reflected at different points in the IoT product life cycle are key to long term system resilience. Optimal management of legacy systems that may be forgotten, unpatched and original technology vendors have long gone out of business is difficult. Maintaining oversight and updating firmware on distributed industrial IoT systems in a systematic manner will be even harder than the existing logistical challenges faced for in-house IT systems. Furthermore, the temporality of organisational security practices needs reflection, as management changes, processes are less well enforced, assets are hired, sold or decommissioned (perhaps even to competitors).

**Implementation Gap for Best Practice** - In guarding against these threats, finding best practices to secure systems is critical and whilst guidelines[172] might be emerging, implementation must catch up. In practice, it is likely there will be a period of coexistence between legacy systems and new IoT devices, as we see in the domestic IoT. Furthermore, skills gaps for employees could be a key vulnerability and securing IoT infrastructure requires creating systems that are usable for workers, and retraining to ensure they are used correctly.

**Managing Infrastructural Complexity** - Systematic approaches considering how best to build security into these systems need to contend with the interdependent, complex nature of industrial systems (e.g. energy, manufacturing, logistics). Even if one element of a system puts appropriate security safeguards in place, when interacting with systems lacking these, vulnerabilities can surface. Over the course of IoT system life cycles, flaws will emerge, but the complex interactions between IoT systems may complicate meaningful anticipation of any second order effects. Responsibility for maintaining oversight of security within systems may be tractable, but establishing responsibility for the points where they interact with other systems may be harder. However, these challenges need to be addressed to avoid the emergence of the internet of insecure industrial things.

**Acknowledgements:**
This work was supported by the Engineering and Physical Sciences Research Council [grant number EP/M02315X/1].

**Abbreviations:**
APTs = Advanced Persistent Threats
CPS = Cyber Physical Systems
C&Cs = Command and Control servers

---

[172] See maintained list of IoT Security and Privacy Guidelines on Schneier on Security Blog https://www.schneier.com/blog/archives/2017/02/security_and_pr.html

CMA = UK Computer Misuse Act 1990
DDoS = Distributed Denial of Service
GDPR = General Data Protection Regulation 2016
GPS = Global Positioning System
ICS = Industrial Control Systems
IoT = Internet of Things
IIoT = Industrial Internet of Things
NIS = Network and Information Security
RFID = Radio Frequency Identification